\documentclass[letterpaper, 12pt]{article}
\usepackage{latexsym}
\usepackage{amsmath}
\usepackage{amssymb}
\usepackage[latin1]{inputenc}
\usepackage{graphicx}
\newcommand{\U}{\textrm{U}}

\def\CN{{\cal N}}

\newcommand{\ads}[1]{{\rm AdS}_{#1}}

\newcommand{\be}{\begin{equation}}
\newcommand{\ee}{\end{equation}}
\newcommand{\bea}{\begin{eqnarray}}
\newcommand{\eea}{\end{eqnarray}}

\newcommand{\labell}[1]{\label{#1}\qquad_{#1}} 
\newcommand{\bbibitem}[1]{\bibitem{#1}\marginpar{#1}}
\newcommand{\llabel}[1]{\label{#1}\marginpar{#1}}

\def\Label#1{\label{#1}%
  \smash{\hbox to0pt{\raise1ex\hbox{\tiny[#1]}\hss}}}
\def\noLabels{\let\Label=\label}
\def\nolabells{\let\labell=\label}
\def\nollabels{\let\llabel=\label}
\def\nobbibitem{\let\bbibitem=\bibitem}

\title{{\bf Quantum geometry and gravitational entropy}}

\author{Vijay
Balasubramanian$^\sharp$, Bart{\l}omiej Czech$^\sharp$, Klaus
Larjo$^\sharp$, \\
Donald Marolf$^\dagger$ and Joan Sim\'{o}n$^\flat$
\footnote{vijay@physics.upenn.edu, czech@sas.upenn.edu,
klarjo@physics.upenn.edu, marolf@physics.ucsb.edu,
JSimonSoler@lbl.gov}
\\[1mm]
\small \sl $^\sharp$\; David Rittenhouse Laboratories, University of Pennsylvania,
\\[-1.5mm]
\small \sl Philadelphia, PA 19104, USA \\
\small \sl $^\dagger$Physics Department, UCSB,\\
[-1.5mm]
\small \sl Santa Barbara, CA 93106, USA\\
\small \sl $^\flat$ \; Dept of Physics \& Theoretical Physics Group, LBNL,
\\[-1.5mm]
\small \sl University of California, Berkeley,  CA 94720, USA  \\
}

\begin{document}

\noLabels \nolabells \nollabels \nobbibitem
\setlength{\baselineskip}{16pt}
\begin{titlepage}

\maketitle

\vspace{-4in}
\rightline{\small UPR-T-1182}
\rightline{\small UCB-PTH-07/09}
\rightline{\small hep-th/0705.4431}
\vspace{4in}

\begin{abstract}
Most quantum states have wavefunctions that are
widely spread over the accessible Hilbert space and hence do not
have a good description in terms of a single classical geometry.
In order to understand when geometric descriptions are possible,
we exploit the AdS/CFT correspondence in the half-BPS sector of
asymptotically $\ads{5} \times S^5$ universes. In this sector we
devise a ``coarse-grained metric operator" whose eigenstates are
well described by a single spacetime topology and geometry.
We show that such half-BPS universes have a
non-vanishing entropy if and only if the metric is singular, and that the
entropy arises from coarse-graining the geometry. Finally, we use our entropy
formula to find the most entropic spacetimes with fixed
asymptotic moments beyond the global charges.
\end{abstract}
\thispagestyle{empty} \setcounter{page}{0}
\end{titlepage}


\tableofcontents

\section{Introduction}
\llabel{intro} The $\frac{1}{2}$-BPS sector of $\mathcal{N} = 4$
$\textrm{SU}(N)$ Yang--Mills theory provides a simple playground
for understanding quantum gravity, as the semiclassical mapping
between spacetime geometries and coherent field theoretic states
is particularly simple
\cite{llm,BBNS,cjr,berenstein,vijayjoan,janetal,larjo}.   Here we
propose an extension of this mapping to the quantum level.  Using
a second-quantized formalism, we define a `metric' operator  in
the Yang-Mills theory whose eigenstates map to universes with a
single well-defined topology and geometry, while non-eigenstates
do not have a good description in terms of single spacetimes.

In the field theory, the half-BPS states can be constructed by
reducing the $\textrm{SU}(N)$ Yang-Mills Lagrangian to a Hermitian
matrix model and studying its Hilbert space \cite{cjr,berenstein}.
This in turn has  a description in terms of  $N$ fermions in a
harmonic potential.  Approximate eigenstates of our metric
operator can be constructed by placing the individual fermions in
coherent states.  (See \cite{janetal,larjo} for a related
discussion in the $\ads{3}$ context.)  Using this formalism, we
can associate an entropy to any classical, asymptotically $\ads{5}
\times \textrm{S}^5$ spacetime in the LLM family given by
\cite{llm}, by counting how many microscopic configurations give
the same coarse-grained metric.  We find that only the singular
spacetimes have a finite entropy; thus spacetime singularities and
the appearance of gravitational entropy are intimately tied
together, at least within the class of LLM geometries.     By
studying the partition sum over geometries we show that only the
smooth spacetimes appear in the underlying configuration space,
which is thus effectively discretized.  This provides further
evidence that singularities and their associated entropy arise
only in passing to our long-distance effective description. Our
formalism can be used to determine the most entropic geometries
with specified angular moments; in essence, this extends the
extremal ``superstar'' black hole geometry to black objects
carrying higher angular moments.

The data required for constructing the metric in these half-BPS
universes can be extracted from the effective single-particle
phase space distribution of the fermionic representation of the
half-BPS matrix model.   This procedure generically loses
information about the complete state, which is fully represented
in an  $N$-particle phase space.   Hence the semiclassical
geometry will generally lose information about the underlying
quantum state.  We show that in the semiclassical limit ($N \to
\infty$)  this sort of  information loss occurs if and only if the
metric is singular.

While this paper was in preparation we received \cite{derrico}
which has substantial overlap with some sections of this paper,
and \cite{marketal} which carries out similar analyses of the
entropy of coarse-grained BPS spacetimes  in other settings.

\section{Half-BPS states: review and set up}
\Label{HalfBPSstates}

\paragraph{Gravity: }
Type IIB string theory in 10 dimensions has a family of
asymptotically $\ads{5} \times S^5$ solutions that preserve 16
supercharges; these are the $\frac{1}{2}$-BPS states. Of
particular interest are the subset of states that preserve an
additional $\textrm{SO}(4) \times \textrm{SO}(4)$ symmetry. All
states in this subset were written down in \cite{llm} and are of
the form
\begin{equation}
ds^2 = -h^{-2} \, (dt+V_idx^i)^2 + h^2 \, (dy^2 + dx^idx^i) + R^2 \, d\Omega_3^2 + \tilde{R}^2 \, d\tilde{\Omega}_3^2,
\Label{llmsolution}
\end{equation}
where the coefficients are given by
\begin{equation}
R^2 = y \, \sqrt{\frac{1-u}{u}}, \quad \tilde{R}^2 = y \, \sqrt{\frac{u}{1-u}}, \quad h^{-2} = \frac{y}{\sqrt{u(1-u)}},
\Label{llmfunctions1}
\end{equation}
and are entirely specified by the function $u(x_1,x_2,y)$. The
one-form field $V$ can also be given in terms of $u$, but it will
not be needed here.  Note that $x_1$ and $x_2$ have units of
length squared.  This function in turn solves a harmonic equation
in $y$ and is  determined by its boundary condition on the $y=0$
plane
\begin{equation}
u(x_1,x_2) \equiv u(x_1,x_2,0) \, . \Label{udef}
\end{equation}
(We will frequently drop the arguments and simply call this
function $u$.) On this space of solutions it can be shown
\cite{llm} that the Hamiltonian is
\begin{equation}
H = \int {d^2 x \over 2\pi \hbar} {x_1^2 + x_2^2 \over 2 \hbar} \, u(x_1,x_2)  \, . \Label{Hdef}
\end{equation}
The flux of $F_5$ passing through the asymptotic $S^5$ in such spaces is
\begin{equation}
N = \int {d^2 x \over 2\pi \hbar} \,  u(x_1,x_2) \Label{Ndef}
\end{equation}
It turns out that the solutions (\ref{llmsolution}) are singular
unless $u(x_1,x_2) = 0,1$ everywhere.   When $u >1$ or $u<0$ the
solutions have closed timelike curves \cite{milanesi}, so we will
always take $0\leq u \leq 1$.

\paragraph{Field theory: }
In the dual $\CN = 4$ $\textrm{SU}(N)$ Yang-Mills theory on $S^3
\times R$ the corresponding states preserve 16 supersymmetries.
Due to the BPS condition these states are constant on the
$\textrm{S}^3$, leading to an $\textrm{SO}(4)$ symmetry. When one
constructs states using only one of the three complex scalar
fields, the internal $\textrm{SO}(6)$ R-symmetry is broken to
$\textrm{SO}(4)$. Thus these states have the same symmetries as
the gravity solutions described above.

It has been shown \cite{cjr,berenstein} that all of these states
can be described in terms of a matrix model for the one homogenous
mode of the complex adjoint scalar field:
\begin{equation}
S_{{\rm YM}} = {1 \over g^2_{YM}} \int dt \left( {1 \over 2} \dot{X}^2 + {1 \over 2} X^2 \right) \, . \Label{YMaction}
\end{equation}
The dictionary relating that Yang-Mills theory to string theory ensures that
\begin{equation}
\hbar \leftrightarrow \ell_P^4 \Label{hbarellp}
\end{equation}
where $\hbar$ is Planck's constant governing the semiclassical
limit of (\ref{YMaction}) and $\ell_P$ is the 10 dimensional
Planck length governing the semiclassical limit on the gravity
side.   The matrix model (\ref{YMaction}) can be solved in the
usual way by going to the eigenvalue basis, in which case it
reduces to a system of $N$ non-interacting fermions in a harmonic
potential. A basis for the Hilbert space is specified by a set of
non-decreasing integers
\begin{equation}
r_i = \frac{E_i}{\hbar} - i + 1/2,~~~ i=1,2,\cdots N \Label{rdef}
\end{equation}
which measure the excitation of each fermion above the vacuum. The
data $\{r_i\}$ may be summarized in a Young diagram in which the
${\rm i}^{\rm th}$ row has length $r_i$. In the semiclassical
limit, it has been proposed that these states are mapped to the
dual solutions (\ref{llmsolution}) with the
identification\footnote{Here we reabsorb a factor of $2\pi$ into
the definition of $W(p,q)$ in comparison with \cite{vijayjoan} in
order to maintain consistency with Sec.~4.2 of \cite{wigner}.}
\cite{vijayjoan}
\begin{equation}
u(x_1,x_2) \leftrightarrow \hbar\, W(p,q) ~;~~~~~~~ (x_1,x_2)
\leftrightarrow (p,q) \Label{uWcorrespond}
\end{equation}
where $W(p,q)$ is the semiclassical density of fermions in the
single particle harmonic oscillator phase plane $(p,q)$.   There
are many different quantum mechanical density functions on phase
space; however, all well-defined phase space distributions share
the same semiclassical limit.   Thus we will mostly work with the
well-known Wigner density function (see Appendix A) and
occasionally with the Husimi distribution (see
Sec.~\ref{husimisec}) that is obtained by convolving the Wigner
distribution with a coherent state \cite{vijayjoan}.

\paragraph{Integrable charges: }

In \cite{us} we pointed out that the half-BPS sector of $\CN=4$
Yang-Mills theory is integrable and thus has a family of higher
Hamiltonians
\begin{equation}
M_k = {\rm Tr}(H_N^k/\hbar^k) = \sum_{i=1}^N \lambda_i^k  ~~~~;~~~~ k = 0, \cdots N
\Label{higherham}
\end{equation}
where $H_N$ is the Hamiltonian in the free fermion Hilbert space
and $\lambda_i = \frac{E_i}{\hbar}$ is the energy of the $i^{th}$
fermion in units of $\hbar$.  This can be written in terms of a
phase space integral, and using the correspondence with spacetime
(\ref{uWcorrespond}), becomes
\begin{equation}
M_k = \int \frac{d^2x}{2\pi \hbar} \frac{(x_1^2+x_2^2)^{k}}{2^k\hbar^k} u(x_1,x_2), \Label{Mk}
\end{equation}
which is accurate to leading order in $N$.

Any eigenstate of this tower of conserved charges can be
completely identified by measurement of all the eigenvalues.    In
\cite{us} we showed that in the semiclassical limit these higher
Hamiltonians can be read off from the asymptotic angular moments
of the spacetime.   Specifically, the k$^{{\rm th}}$ angular
moment of the function $u(x_1,x_2,y)$ that appears in
(\ref{llmsolution}) is proportional to the expected value of
$M_k$:
\begin{equation}
u(\rho,\theta,\varphi) = 2\cos^2 \theta \sum_{k=0}^{\infty} \frac{2^k \langle M_k\rangle }{\rho^{2k+2}} \, (-1)^k(k+1)\, \,
{_2F_1}(-k,k+2,1;\sin^2 \theta). \Label{eq:wignerlimit}
\end{equation}
where $\rho^2 = y^2 + r^2$, $y$ is the transverse direction to the
$(x_1,x_2)$ plane, $r$ is the radial direction in this plane and
$\theta$ is the angle between the plane and the transverse
direction measured by $\rho$.\footnote{The function $u$ is
independent of $\phi$ because we are dealing with energy
eigenstates.} Nevertheless, it turns out that in typical states
with a given total energy ($M_1$), Planck scale measurements are
required to distinguish between the microstates.   Hence
information about these states is lost in the semiclassical limit.

\section{Quantum geometry}
\llabel{quantum}

The naive correspondence  (\ref{uWcorrespond})  associates a
geometry to every state in the Hilbert space via the phase space
fermion density in the free fermi description of the half-BPS
states.    The correspondence cannot be correct in this naive form
-- while a suitably coherent state in the field theory can
correspond to a semiclassical geometry, there will be many states
which correspond to wavefunctions that have support on spacetimes
of widely different geometry and topology.      Thus it is of
interest to identify which field theory states have a description
in terms of a single classical geometry vs. a superposition of
geometries.    In this section, we will define a `metric' operator
in the Yang-Mills theory and propose that the eigenstates of this
operator are the ones that can be mapped to semiclassical
geometries using (\ref{uWcorrespond}), while the non-eigenstates
cannot be associated to a unique metric in this manner.

\subsection{Four ways of becoming a superstar?}
\llabel{superstar-example}

\begin{figure}[t]
\begin{center}
\includegraphics[scale=0.4]{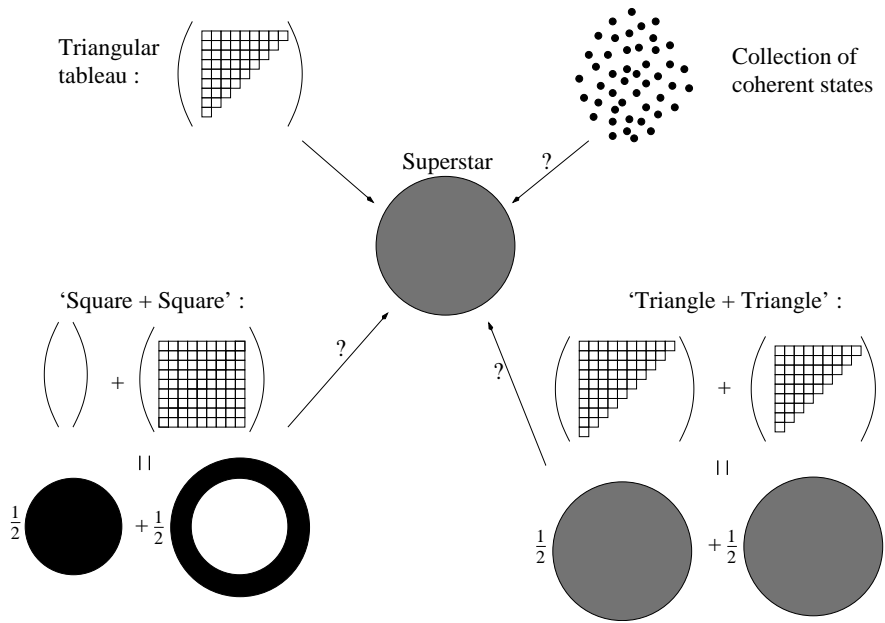}
\end{center}
\caption{\label{superstars} Four CFT states naively would be mapped to the superstar.}
\end{figure}

To illustrate the difficulty with the naive correspondence
(\ref{uWcorrespond}) we will consider different ways in which the
extremal half-BPS black hole (the ``superstar''
\cite{myerstafjord}) would be generated by applying
(\ref{uWcorrespond}) to field theory states.      It was shown in
\cite{llm,vijayjoan} that the metric (\ref{llmsolution}) realizes
the superstar geometry when the function $u(x_1,x_2)$ (\ref{udef})
takes a constant value between 0 and 1 within a circular droplet
in the $(x_1,x_2)$ plane, and vanishes outside it.
This $u$ function is depicted in the center of
Fig.~\ref{superstars}.   It was also shown that a fermion basis
state (\ref{rdef}) described by a triangular Young tableau (upper
left corner of Fig.~\ref{superstars}) has a phase space density of
this form, and hence corresponds to a superstar geometry according
to (\ref{uWcorrespond}).

We will now examine other states in the field theory that have the
same single particle Wigner distribution.  For definiteness, we choose, without
loss of generality, to work with a droplet with $u(x_1,x_2) =
\frac{1}{2}$ inside the droplet. In terms of the triangular
tableau, this corresponds to having an equal number of rows and
columns ($N = N_c$) \cite{vijayjoan}, while for the superstar this
means that the number of giant gravitons sourcing the geometry is
equal to the flux of the five-form field \cite{myerstafjord}. The
radius of the droplet for this configuration is
$\sqrt{2}R_{AdS}^2$.\footnote{Remember that the dimensions of
$x_1$ and $x_2$ (and therefore $r$) are (length)${}^2$.}   In
terms of excitation numbers the triangular tableau state is given
by
\begin{equation}
| \textrm{`triangle'} \rangle = |0,2,4,\ldots, 2(N-1) \rangle.
\end{equation}

Another way of producing the superstar is to place the $N$
fermions in the field theory in coherent states that are randomly
distributed in  phase space inside the circle of radius
$\sqrt{2}R_{AdS}^2$ (upper right corner, Fig.~\ref{superstars}).
Since coherent states are minimum uncertainty packets, as we will
see in Secs.~\ref{coarsegrained} and \ref{partition}, they will
cover half the area contained inside the circle.  Upon coarse
graining at the $\hbar$ scale, this configuration has a phase
space distribution that is $\hbar W = 1/2$ inside the circle and
$\hbar W=0$ outside.  Upon using the correspondence
(\ref{uWcorrespond}) we once again recover the superstar geometry.

Thirdly, we can consider a superposition of two very similar
triangles, one with even excitations and one with odd excitations
(Fig.~\ref{superstars};  bottom right):
\begin{eqnarray}
| \textrm{\textrm{`triangle + triangle'}} \rangle &=& \frac{1}{\sqrt{2}} |0,2,\ldots,2N-2 \rangle + \frac{1}{\sqrt{2}} |1,3,\ldots,2N-1\rangle \nonumber \\
 & \equiv & \frac{1}{\sqrt{2}} | \textrm{`evens'} \rangle + \frac{1}{\sqrt{2}} | \textrm{`odds`} \rangle.
\Label{triangletriangle}
\end{eqnarray}
And finally, we can consider the superposition of the empty
tableau corresponding to the $AdS$ vacuum, and a square tableau of
size $N_c \times N$ (Fig.~\ref{superstars};  bottom left):
\begin{eqnarray}
| \textrm{\textrm{`square + square'}} \rangle &=& \frac{1}{\sqrt{2}} |0,1,\ldots,N-1\rangle + \frac{1}{\sqrt{2}}
|N,N+1,\ldots,2N-1\rangle \nonumber \\
& \equiv & \frac{1}{\sqrt{2}} | 0 \rangle + \frac{1}{\sqrt{2}} | N \times N_c \rangle.
\Label{squaresquare}
\end{eqnarray}
The last two superpositions are also depicted in figure
\ref{superstars}. Computing the corresponding phase space
distributions (following \cite{vijayjoan} or (\ref{eq:wigner}) later in this paper)
one can see that the phase space distributions corresponding to these states will indeed reproduce
the superstar geometry upon coarse graining.

The last two cases are both simple superpositions of basis states.
However, the `square + square' example is more dramatic as it is a
superposition of states of vastly different energies, unlike
`triangle + triangle'. The energy associated to the $AdS$ vacuum
is $\frac{N^2}{2}$, while the energy of the $N \times N_C$ tableau
is $\frac{3N^2}{2}$ for $N = N_c$, yielding an average energy of
$N^2$.  As soon as an observer performs even a rough measurement
of the energy,  the universe is projected into one of the two
energy eigenstates. On the gravity side, the corresponding
geometries differ from each other (and from the superstar!) at
scales that are easily measurable, since the relevant scale is
roughly the $AdS$ radius $R_{AdS}$. Thus, `square + square' does
not describe anything like a classical superstar, and can only be
described as a superposition of distinct geometries.   We wish to
develop a formalism that can distinguish states of this sort from
the ones that can be usefully mapped to a unique metric.

\subsection{The second quantized formalism}
\Label{2nd}
It will be convenient to use a second quantized
description of the fermion system.  For each state $\psi$ in the
1-particle Hilbert space ${\cal H}_{1}$, we define both a fermion
creation operator $b^\dagger(\psi)$ and a fermion annihilation
operator $b(\psi)$ which satisfy

\begin{eqnarray}
\{ b(\psi_1), b(\psi_2) \} =0, \cr
 \{ b^\dagger (\psi_1),
b^\dagger(\psi_2) \} =0, \cr
 \{ b(\psi_1), b^\dagger(\psi_2) \} = \langle \psi_1 | \psi_2
 \rangle, \Label{acom}
 \end{eqnarray}
where the inner product on the right is taken in ${\cal H}_{1}$.
Since we deal with free fermions, the anti-commutators on the
left-hand side of equations (\ref{acom}) are proportional to the
identity.  Thus it makes sense to set them equal to the complex
numbers on the right-hand side.

We will in particular be interested in the case where $\psi$ is a
coherent state labelled by a parameter $\alpha \in {\mathbb C}$.
In phase space these states manifest themselves as gaussian
wavepackets localized around $\alpha =
\frac{x_1+i\,x_2}{\sqrt{2\hbar}}$, and are defined by
\begin{equation}
|\alpha \rangle = e^{-|\alpha|^2/2} \sum_{n=0}^\infty \frac{\alpha^n}{\sqrt{n!}}|n\rangle \equiv \sum_{n=0}^\infty c_n(\alpha)
|n\rangle\, . \Label{eq:coherent}
\end{equation}
These form an overcomplete basis, and have overlaps
\begin{equation}
\label{overlap} \langle \beta | \alpha \rangle =
\exp\left(-\frac{|\alpha|^2+|\beta|^2-2\beta^* \alpha}{2}\right) \, .
\end{equation}
Throughout this section, the normalization conventions are as in
\cite{wigner}. We shall abuse notation somewhat by taking
$b^\dagger(\alpha), b(\alpha)$ to represent the creation and
annihilation operators corresponding to the coherent state
determined by  $\alpha \in {\mathbb C}$.  The simple form and
well-known properties of the coherent state (\ref{eq:coherent})
provide a convenient foundation upon which to base a semiclassical
formalism.  We make use of this opportunity in the rest of this
work.  However, in section \ref{discussion}, we describe certain
desirable improvements  which, unfortunately, will come with
increased technical complications.

We wish to define a `metric' operator in the field theory with the
property that its eigenstates are well described by a
single classical metric which is in turn given by the eigenvalue
of the operator. Since any 1/2-BPS metric may be expressed in
terms of the function $u(x_1,x_2)$, it is enough to define an
operator $\hat u(x_1,x_2)$ in terms of the fermion operators; we
can take all other metric operators to be given by the classical
LLM formulae with $u$ replaced by $\hat u$. We will see that,
with our definition of $\hat u$ below, such expressions will
require no regularization.

For $\alpha = \frac{x_1 + i x_2}{\sqrt{2\hbar}}$, we make the
definition

\begin{equation}
\hat u(\alpha) \equiv b^\dagger(\alpha) b(\alpha),
\end{equation}
and $\hat u(x_1,x_2) \equiv \hat u (\frac{x_1 + i
x_2}{\sqrt{2\hbar}})$. Thus, for each $\alpha$ the operator $\hat
u(\alpha)$ is a fermion number operator, having eigenvalues 0 and
1.

Our definition satisfies a number of useful properties.  For example, we have
\begin{eqnarray}
 \hat{N} & = & \int \frac{dx_1 dx_2}{2\pi \hbar}\, \hat u(x_1,x_2), \
\ \ {\rm and} \\  \hat{H} & = & \int \frac{dx_1 dx_2}{2\pi \hbar} \frac{x_1^2 + x_2^2}{2 \hbar} \,\hat u(x_1,x_2), \ \ \ {\rm
and}
\\ \hat{M}_k & = & \int \frac{dx_1 dx_2}{2\pi \hbar} \left( \frac{x_1^2 + x_2^2}{2 \hbar} \right)^k \hat u(x_1,x_2).
\Label{useful}
\end{eqnarray}
These are precisely the relations which hold between the
corresponding classical quantities.

Furthermore, one may show that these $\hat u(\alpha)$ generate a
complete set of operators which do not change fermion number. Here
the essential point is that any operator on the 1-particle Hilbert
space can be expressed in the form $\int d^2 \alpha f(\alpha)
|\alpha \rangle \langle \alpha |$.  Now, if $|\psi_n \rangle$
denotes the ${\rm n}^{\rm th}$ energy level of the 1-particle
harmonic oscillator, we may define a complex function
$f_{nm}(\alpha)$ by
\begin{equation}
|\psi_n \rangle \langle \psi_m | = \int \! \! \! \int d^2 \alpha f_{nm}(\alpha) |\alpha \rangle \langle \alpha |.
\end{equation}
It is then straightforward to show that
\begin{equation}
b^\dagger(\psi_n) b(\psi_m) = \int \! \! \! \int d^2 \alpha f_{nm}(\alpha) \hat u( \alpha).
\end{equation}
It is clear that any operator which preserves particle number can
be built by taking sums of products of $b^\dagger(\psi_n)
b(\psi_m)$, and thus that any such operator can be built from
$\hat u(\alpha)$.

\subsubsection{$\hat{u}$ and the Husimi distribution}
\Label{husimisec}
We now wish to connect the operator $\hat{u}$ to
a semi-classical phase space distribution function by showing that
the expectation value of $\hat{u}$ in a general state
$|\Psi\rangle$ is the Husimi distribution on phase space:
\begin{equation}
\langle \Psi | \hat{u}(\alpha) | \Psi \rangle  = \pi
\textrm{Hu}^{\rho_1}(\alpha), \Label{husimi}
\end{equation}
where $\textrm{Hu}^{\rho_1}(\alpha)$ is the Husimi distribution
corresponding to the one particle density matrix $\hat{\rho}_1$
associated to $|\Psi\rangle$. To show this, we write the general
state as a superposition of basis states as
\begin{equation}
|\Psi \rangle = \sum_w d_w |\mathcal{F}^w \rangle,
\end{equation}
where $w$ sums over the basis states in the superposition, the
normalization is $\sum_w |d_w|^2 =1$ and $\mathcal{F}^w$ is the
set of excitation numbers characterising a basis state
$|\lambda_1^w,\ldots,\lambda_N^w\rangle$. The basis states in turn
are given by
\begin{equation}
|\lambda_1, \ldots, \lambda_N \rangle = b^{\dag}(\lambda_1) \ldots b^{\dag}(\lambda_N) |0\rangle .
\end{equation}
Using the commutators (\ref{acom}) one can show that
\begin{eqnarray}
& & \langle \Psi |  \hat{u} (\alpha) | \Psi \rangle = \sum_{w,w'} d^*_{w'} d_{w} \sum_{i_w,j_{w'}=1}^N (-1)^{i_w+j_{w'}} \langle
\lambda_{j^{w'}}^{w'} | \alpha \rangle \langle \alpha | \lambda_{i^w}^w \rangle  \langle 0 | \prod_{\stackrel{k=1}{k \neq i_w}}^N
b(\lambda_k^{w'}) \prod_{\stackrel{l=1}{l\neq j_{w}}}^N b^{\dag}(\lambda_l^{w}) | 0 \rangle \nonumber \\
& & = \sum_w |d_w|^2 \sum_{i_w=1}^N  |\langle \alpha | \lambda_{i^w}^w \rangle|^2 + \sum_{(w,w')\in \mathcal{C}} d_w d_{w'}^*
(-1)^{N^{(w,w')}+N^{(w',w)}}  \langle \lambda_{(w',w)} | \alpha \rangle \langle \alpha | \lambda_{(w,w')} \rangle .
\Label{husexp}
\end{eqnarray}
The expectation value on the right side of the first line can
clearly be nonzero only when at least $N-1$ of the excitation
numbers in $\mathcal{F}^w$ and $\mathcal{F}^{w'}$ are the same.
This allowed us to simplify the expression at the cost of
introducing some new notation. In the above, $\mathcal{C}$ denotes
the set of pairs of states that differ from each other by one
excitation, and for each such pair we define the excitation to be
$\{ \lambda_{(w,w')} \} = \mathcal{F}^w \backslash
\mathcal{F}^{w'}$. Finally, $N^{(w,w')}$ indexes the number of the
differing excitation, i.e. if $\lambda_{(w,w')} = f_i^w \in
\mathcal{F}^w$, then $N^{(w,w')} = i$.

Next we need to evaluate the right hand side of (\ref{husimi}),
and thus we recall the definition of the Husimi distribution from
\cite{wigner}:
\begin{equation}
\textrm{Hu}^{\rho_1}(\alpha) = \frac{1}{\pi} \langle \alpha | \hat{\rho}_1 | \alpha \rangle. \Label{husdef}
\end{equation}
The one particle density matrix corresponding to $|\Psi \rangle$
can be easily found by tracing over $N-1$ of the particles. Due to
the orthogonality, the trace can also be nonzero only if at least
{\sf $(N-1)$} of the excitations are the same, and after a little
work we get
\begin{eqnarray}
& & \hat{\rho}_1 = N \, \textrm{Tr}_{\mathcal{H}_2\otimes \ldots
\mathcal{H}_N} ( | \Psi \rangle \langle \Psi | ) = \sum_w
|d_w|^2 \sum_{i_w=1}^N  | \lambda_{i^w}^w \rangle \langle \lambda_{i^w}^w | + \nonumber \\
& & + \sum_{(w,w')\in \mathcal{C}} d_w d_{w'}^*(-1)^{N^{(w,w')}+N^{(w',w)}}  | \lambda_{(w,w')} \rangle \langle \lambda_{(w',w)}
| \Label{eq:rho1},
\end{eqnarray}
and plugging this into (\ref{husdef}) we immediately recover
(\ref{husimi}).

\subsection{A coarsegrained $\hat{u}$ and its eigenstates}
\llabel{coarsegrained} The metric operator $\hat{u}$ is not well
suited to a semiclassical observer, since it is very sensitive to
details at the Planck scale. Thus, we wish to coarse grain it over
some distance scale $L$. To do this, we'll compute a convolution
with a Gaussian kernel:
\begin{equation}
\tilde{u}(x_1,x_2) = \frac{1}{\pi L^2} \int \! \! \! \int dx'_1 \,
dx'_2 \, e^{-\frac{(x_1-x'_1)^2+(x_2-x'_2)^2}{L^2}}
\hat{u}(x'_1,x'_2).
\end{equation}
Since this operator is essentially a weighted average of local
values of $\hat{u}$, it will have a continuous spectrum of
eigenvalues between 0 and 1 and it is not sensitive to details at
the Planck scale.  This operator valued function on the single
fermion phase space is  our proposal for a ``metric operator'',
namely the operator whose eigenstates can be associated to
semiclassical geometries using the LLM prescription.

To make this more precise, we next need to define exactly what me
mean by eigenstates and the semiclassical limit. The semiclassical
limit can be defined simply as
\begin{equation}
\ell_P \to 0 ~~~~;~~~~ L \to 0 ~~~~;~~~~ {L \over \ell_P} \to \infty. \Label{semilimit}
\end{equation}

We can see that the commutator $[ \tilde{u}(\alpha),
\tilde{u}(\beta)]$ does not vanish, though it approaches zero
rapidly when $\alpha$ and $\beta$ are separated by more than a
distance $L$.  As a result, only states which are eigenstates of
the $\hat{u}(\alpha)$ with eigenvalue $0$ or $1$ for all $\alpha$
(i.e. states with an empty or completely filled phase plane) can
be exact eigenstates of $\tilde{u}$.  Neither of these
possibilities is of interest. Other states are only approximate
eigenstates.   For example, consider empty AdS space, the vacuum
of the theory.   In terms of Young tableaux for fermion excitation
energies, AdS space is described as the empty tableau and the
correspondence phase space density is the filled fermi sea, i.e. a
filled disk of fermions in phase space.  For $|\alpha|$ well
within the filled disk, the state is an approximate eigenstate of
$\tilde{u}(\alpha)$ with eigenvalue one. When $|\alpha|$ is well
outside the black disk, the state is an approximate eigenstate of
$\tilde{u}(\alpha)$ with eigenvalue zero. However, when $\alpha$
is within $L$ of the boundary, the state is far from being an
eigenstate of $\tilde{u}(\alpha)$.

Thus we will define approximate eigenstates as follows: We say
that a state $| \Psi \rangle$ is an approximate eigenstate of
$\tilde{u}(\alpha)$ with eigenvalue function $u(\alpha)$ and a
given accuracy $\epsilon$, if and only if
\begin{equation}
\int d^2 \alpha \left| \big( \tilde{u}_N(\alpha) - u(\alpha)\big) |\Psi \rangle \right|^2 < \epsilon.
\end{equation}
With this definition we can easily see that of the four aspiring
superstars introduced earlier, only `square + square' is not an
approximate eigenstate of $\tilde{u}$ thus shouldn't be associated
with the superstar geometry.

Armed with the definition of eigenstates of the coarse-grained
$\tilde u$, such eigenstates can be associated with a
slowly-varying density function which we have called
$u(\alpha)$, and it is in terms of this function that we shall now
discuss the thermodynamics and entropy of these solutions.

\section{Entropy of BPS geometries}
\Label{partition}

Following Sec.~\ref{2nd} all eigenstates of the metric operator
can be constructed by placing each fermion in the free-fermi
description of half-BPS states in a coherent state in phase space.
It is known that coherent states placed on an $\hbar$-spaced
lattice in phase space also provide a complete basis for the
Wigner distributions in the single-particle phase space
\cite{coherentstates}, and hence, by (\ref{uWcorrespond}) for all
semiclassical geometries.   Because of the coarse-graining, many
different quantum states may have the same geometrical
description.   Thus some geometries should be associated to an
entropy, whose exponential counts the number of underlying states
with the same geometric description.    In this section we want to
find a formula for the entropy of half-BPS spacetimes and then use
it to discuss the partition sum over these geometries.  With this
motivation we will construct the Wigner distribution of the
coherent state basis for half-BPS states and coarse-grain that to
find our formula for the entropy of half-BPS semiclassical
spacetimes.

\subsection{Coherent states and entropy}
\Label{sec:coherent}

The semiclassical limit of half-BPS states in Yang-Mills theory is
most conveniently studied in terms of coherent states
(\ref{eq:coherent}), rather than the energy eigenstates described
above.     Each of the $N$ fermions in the harmonic potential can be
placed separately in  a coherent state, with the fermionic
statistics imposed by a Slater determinant.

Consider an $N$-particle (unnormalized) wavefunction given by
antisymmetrizing 1-particle wavefunctions $\Psi_i(x_j)$, $i,j=1
\ldots N$:
\begin{equation}
\Psi(x_1,\ldots,x_N) = \det \left[ \begin{array}{cccc}
      \Psi_1(x_1) & \Psi_1(x_2) & \cdots & \Psi_1(x_N)
   \\ \Psi_2(x_1) & \Psi_2(x_2) & \cdots & \Psi_2(x_N)
   \\ \vdots & & &
   \\ \Psi_N(x_1) & \Psi_N(x_2) & \cdots & \Psi_N(x_N) \end{array}
   \right] = \sum_{\sigma \in S_N} (-)^\sigma \prod_i^N
   \Psi_{\sigma(i)}(x_i) \Label{slater}.
\end{equation}
The 1-particle-reduced density matrix of this state is
\begin{equation}
\hat{\rho}_1 = \frac{N \int dx_2 \ldots dx_N  | \Psi \rangle
\langle \Psi |}{\langle \Psi | \Psi \rangle} = \sum_{mn} | \Psi_m
\rangle \rho_{mn} \langle \Psi_n | \, .
\Label{cohdefrho1}
\end{equation}
In terms of the quantities
\begin{eqnarray}
\hat{D}(\xi) & = & e^{\xi \hat{\alpha}^{\dagger} - \xi^*
\hat{\alpha}} \label{cohdef3} \,  ,\\
\chi(\xi) & = & {\rm tr} (\hat{\rho}_1 \hat{D}(\xi) ) \label{cohdef2} \, ,
\end{eqnarray}
the Wigner phase space distribution of $|\Psi \rangle$ is given by
\begin{equation}
\label{cohdef1} W(\alpha, \alpha^*) = \frac{1}{\pi\hbar} \int d^2
\xi\, e^{\alpha \xi^* - \alpha^* \xi} \chi(\xi) \,
\end{equation}
where $\alpha = \frac{p+ i q}{\sqrt{2\hbar}}$.  Finally,  if we
define the overlap matrix
\begin{equation}
S_{mn} = \langle \Psi_m | \Psi_n \rangle,
\end{equation}
then the 1-particle reduced density matrix takes the simple form
\begin{equation}
\label{cohdefrho2} \rho_{mn} = \frac{N \sum^{'} (-)^\sigma
\prod_{m \neq i = 1}^N S_{i \, \sigma(i)} }{\sum_{\sigma \in S_N}
(-)^\sigma \prod_i^N S_{i \, \sigma(i)} } = S_{nm}^{-1} ,
\end{equation}
where $\sum^{'}$ goes over $\sigma \in S_N$ subject to
$\sigma(m)=n$.

We are interested in the case where each $|\Psi_n\rangle$ is a
coherent state, namely $|\Psi_n \rangle = |\alpha_n \rangle$.
Then, applying equations (\ref{overlap}-\ref{cohdefrho2}), we get:
\begin{equation}
W(\alpha, \alpha^*) = {\rm tr} (T S^{-1}) \Label{wignercoh}
\end{equation}
with
\begin{eqnarray}
T_{mn}(\alpha, \alpha^*) & = & \frac{1}{\pi\hbar} \int d^2 \xi\,
e^{\alpha \xi^* - \alpha^* \xi} \langle \alpha_m | e^{\xi
\hat{\alpha}^{\dagger} - \xi^* \hat{\alpha}} | \alpha_n
\rangle \nonumber \\
& = & \frac{2}{\hbar} \langle \alpha_m | \alpha_n \rangle\,
e^{-2(\alpha -\alpha_m)^* (\alpha - \alpha_n)} . \Label{deft}
\end{eqnarray}
According to the semiclassical holographic map (\ref{wignercoh})
is identified with the function $u$ in spacetime as in
(\ref{uWcorrespond}).

In the semiclassical limit, the phase space distribution of
fermions can be described as a droplet or set of droplets in the
phase plane. Coherent states were constructed to be minimum
uncertainty droplets. They occupy an area of $2 \pi \hbar$, the
smallest possible quantum of phase space. This agrees with
(\ref{Ndef}) and can be verified by close packing coherent states
to form a Fermi sea. In our conventions, $N$ closely packed
coherent states form a fermi sea of area $2 \pi \hbar N$, covered
with a Wigner distribution which fluctuates around a mean value
$\bar{W} = 1/\hbar$, consistent with (\ref{uWcorrespond}). In a
similar way a length $\propto \sqrt{\hbar}$ is a minimal allowed
separation between coherent states. Attempts to force the states
closer than this distance cause them to delocalize into rings as
illustrated in the example in Fig.~2.

\begin{figure}[!ht]
\begin{center}
\includegraphics[width=0.9 \textwidth]{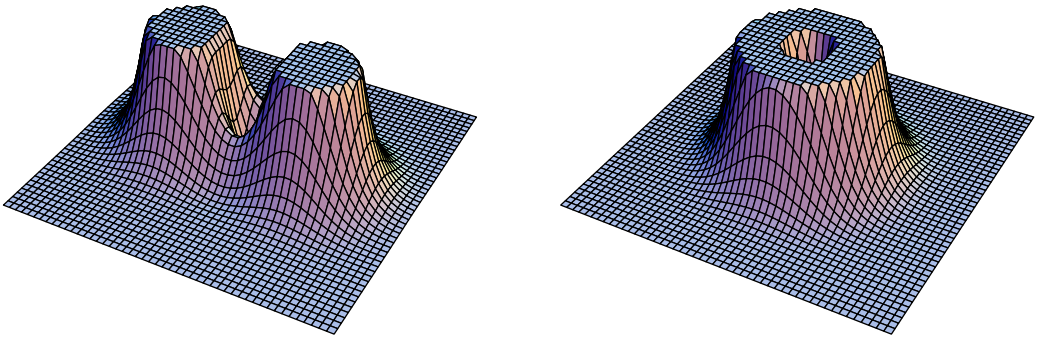}
\end{center}
\caption{Wigner distributions for two coherent states placed at
distances greater than (left) and less than (right) the lattice
size $\propto \sqrt{\hbar}$.}
\end{figure}

The observations above lead to the conclusion that for a
semiclassical observer, it makes sense to think of states as
inhabiting a lattice of unit cell area $2 \pi \hbar$
\cite{coherentstates}.  In fact a semiclassical observer measures
the phase plane at an area scale $\delta A = 2 \pi \hbar M \gg 2
\pi \hbar$. At this scale, the observer is only sensitive to a
smooth, coarse grained Wigner distribution $ 0 \leq \hbar W_c =
u_c \leq 1$ which erases many details of the precise underlying
precise microstates.\footnote{The Wigner distribution can in
general take values greater than 1 or less than 0, but for
coherent states it is always greater than 0. In addition, upon
coarse-graining at a scale bigger that $\hbar$ it lies between 0
and $1 / \hbar$ \cite{vijayjoan}.} We may view the region $\delta
A $ as consisting of $M = \delta A / 2 \pi \hbar$ lattice sites, a
fraction $u_c = \hbar W_c$ of which are occupied by coherent
states. Then the entropy of the local region $\delta A$ is
\begin{equation}
S_K = \log{\binom{M}{\hbar \, W_c \, M}} \sim -M \log{(\hbar
W_c)^{\hbar W_c} (1-\hbar W_c)^{1-\hbar W_c}} = -\frac{\delta
A}{2\pi\hbar} \log{u_c^{u_c} (1-u_c)^{1-u_c}} \, .
\Label{localent}
\end{equation}
The Stirling approximation used in (\ref{localent}) is valid when
$\hbar W_c$ is reasonably far from 0 and 1.   For the total
entropy this gives
\begin{eqnarray}
S & = & \int dS = \int dA \, (\frac{dS}{dA}) \Label{eq:entropy} \\
\frac{dS}{dA} & = & -\frac{u_c \log{u_c} + (1-u_c)
\log{(1-u_c)}}{2\pi\hbar} \, . \Label{entropyequation}
\end{eqnarray}
It is beautiful that thinking about $u_c = \hbar W_c$ as the probability of occupation of a site by a coherent state, this is
simply Shannon's formula for information in a probability distribution.\footnote{Indeed, the same expression for entropy was
arrived at by Masaki Shigemori in an unpublished work by considering a gas of fermionic particles in phase space.}

These facts imply that in the semiclassical limit the function
$u(x_1,x_2)$ which completely defines a classical solution should
effectively be defined on a lattice with each plaquette of area
$\mathcal{O}(\hbar) \leftrightarrow \mathcal{O}(\ell_P^4)$, and
take values of 0 or 1 in each site.  Likewise
(\ref{entropyequation}) will be interpreted as an expression for
the entropy of arbitrary half-BPS asymptotically $\ads{5} \times
S^5$ spacetimes. As an example it exactly reproduces the formula
for the entropy of the typical states described in
\cite{vijayjoan} that correspond in spacetime to the ``superstar''
geometry \cite{myerstafjord}.

Note that the entropy vanishes if and only if $u$ equals $0$ or
$1$ everywhere.   Following the correspondence
(\ref{uWcorrespond}) such states map into geometries that are
non-singular.   We learn that semiclassical half-BPS geometries
that are smooth all have vanishing entropy; and the  presence of
singularities $0 < u < 1$ also implies that the spacetime carries
an entropy.  Thus, in this setting, entropy is a measure of
ignorance of a part of the underlying state which is captured in
classical gravity as a spacetime singularity.

\subsection{The partition function}

In the semiclassical limit the partition function over the half-BPS sector of IIB string theory (${\cal M}$)  with asymptotically
$\ads{5} \times S^5$ boundary conditions can be written as
\begin{equation}
Z = \int_{{\cal M}} \mathcal{D}g_{\mu\nu} \cdots \, e^{-\beta (H(g_{\mu\nu},\cdots) -\gamma N(g_{\mu\nu},\cdots))} =  \int
\mathcal{D} u(x_1,x_2) \, \mu(u)  \,   e^{-\beta (H(u) - \mu N(u) )} \, .
\end{equation}
We are able to write this as a functional integral over just $u$
because the entire classical solution can be derived from this
function, as can the Hamiltonian $H$ and the number of units of
5-form flux $N$.   The measure $\mu(u)$ reflects not only the
Jacobian in transforming between the supergravity fields and $u$,
but also the number of underlying microscopic configurations that
give rise to the same macroscopic spacetime. To derive the
semiclassical measure $\mu$ recall first that the classical metric
is a useful concept at scales $L \gg \ell_P$ and the semiclassical
limit can be defined as (\ref{semilimit}). In this limit the
smooth function $u(x_1,x_2)$ arises by considering the limit of
piecewise constant functions in the lattice of scale $L$.   To
derive the continuum measure $\mu(u)$ we first ask  how many
microscopic configurations can give rise to a given average value
of $u$ in each $L \times L$ lattice cell. From the previous
section we have learned that we can think of $u$ as the average
number of Planck size cells that are populated by coherent states
within an area of size $L \times L$. Following (\ref{localent})
this means that a coarse-grained plaquette of size $L$ with a
given value of $u$ arises from
\begin{equation}
 \binom{L/\ell_P}{u \, L/\ell_P}
 \end{equation}
underlying configurations. Many of these configurations,
representing different occupation numbers of Planck cells, will
have different energies because the energy contribution of a given
Planck cell depends upon its location as $E = (x_1^2 + x_2^2)/2$
(\ref{Hdef}).    However, because $L \to 0$ in the semiclassical
limit, configurations that populate a fixed fraction of Planck
sized cells will necessarily have very similar energies.   In fact
it can be shown that the energies are even similar in Planck units
in the strict semiclassical limit (\ref{semilimit}).   This is
because as $L/\ell_P \to \infty$, the number of Planck cells in
each plaquette is extremely large.    One can use this to show
that the energies of configurations that populate a fixed fraction
$u$ of the Planck cells have a standard deviation to mean ratio
that vanishes as $L/\ell_P \to \infty$.

Putting everything together, the measure on semiclassical half-BPS spacetimes is
\begin{equation}
\mu(u) = e^{-\int {dx_1 \, dx_2 \over 2\pi \hbar} \, \left(u \ln{u}  + (1-u)\ln(1-u)\right) }= e^{S(u)} \Label{measure}
\end{equation}
where $S$ is understood as the entropy of the  spacetime.   In the
semiclassical limit, a spacetime is nonsingular if $u = 0,1$
everywhere.   In that case, $S(\mu) = 0$ and the measure $\mu$ is
1.   In other words, semiclassical half-BPS spacetimes have an
entropy if and only if they are singular.

Defining
\begin{equation}
 f(x_1,x_2) = \beta\left( {x_1^2 + x_2^2 \over 4 \pi \hbar} - {\mu \over 2\pi \hbar}  \right)
\end{equation}
the partition function over semiclassical half-BPS spacetimes becomes
\begin{equation}
Z = \int {\mathcal D}u(x_1,x_2) \,
 e^{-\int {dx_1 \, dx_2 \over 2\pi \hbar} \, \left[ u \ln{u}  + (1-u)\ln(1-u) + f \, u \right] }
 \end{equation}
 Evaluating this by the method of saddlepoints gives
 \begin{equation}
 \ln Z =
 \int \frac{d^2x}{2\pi \hbar} \ln (1 +
e^{-\beta \frac{x_1^2+x_2^2}{2\hbar}+\beta \mu}) = \int_0^{\infty} ds \, \frac{s}{e^{s-\beta\mu}+1} \equiv \frac{1}{\beta}
F_2(e^{\beta\mu}), \Label{partfn}
\end{equation}
where $s= \beta (x_1^2 + x_2^2)/2$ and $F_2$ is a Fermi-Dirac
function.  It is worth emphasizing that we have summed over both
singular and non-singular semiclassical spacetimes, but have
included the correct degeneracy factor (\ref{measure}).  A naive
approach to summing over geometries would have failed to include
this measure because there is no semiclassical horizon giving
arise to a Bekenstein-Hawking entropy.  In such a a naive approach
the partition function would have been
\begin{equation}
\ln \tilde{Z} = \int \frac{d^2x}{2\pi } \ln \left( \frac{1-e^{(-\beta \frac{x_1^2+x_2^2}{2} + \beta \mu)}}{\beta
\frac{x_1^2+x_2^2}{2} - \beta \mu} \right) = \frac{1}{\beta} \int_0^{\infty} ds \ln \left( \frac{1-e^{-s-\beta \mu}}{s-\beta \mu}
\right).
\end{equation}
This integral diverges at the upper limit.     A naive approach
taking the unit measure over geometries would only reproduce
(\ref{partfn}) if we restricted the partition sum to be over
smooth geometries ($u = 0,1$) with $u$ taking constant values
within elementary cells at the Planck scale.  This would mimic in
geometry the coherent state analysis given above, but its validity
is more doubtful because at the Planck scale the wavefunction over
geometries is always relevant.

The semiclassical partition function \eqref{partfn} should
reproduce the partition function obtained from first principles by
coarse-graining the scale of the fundamental cells in the quantum
mechanical phase space. This procedure is not ambiguous because the free fermion gauge theory description provides us with an honest quantum mechanical phase space. Coarse-graining is defined as a renormalization group transformation in this space, and it is the 1/2 BPS nature of the quantum states that allows us to trust this description as the value of the coupling is turned on to achieve a reliable gravitational description. Consider then a lattice whose cells are $M\times M$ (in Planck
units). From the microscopic point of view, the energy of each
distribution of populated Planck scale cells is different, but as
we argued before, in the limit $M\to\infty$, almost all
distributions cluster close to a certain typical distribution in
the $M \times M$ cell, and thus observers at these scales will
assign the same energy to all of them. In this case, the
coarse-grained function $u$ will take values $0, \frac{1}{M^2},
\frac{2}{M^2}, \ldots, 1$ in the $M \times M$ cells. This can also
be seen from the flux quantization \eqref{Ndef} which is scale
independent.  In other words, discretising phase space and
comparing two lattices of sizes $M\times M$ and $1\times 1$, in
Planck units, we find that
\begin{equation}
N =  \sum_{\{x_1,\,x_2\}} u(x_1,\,x_2) = M^2\,\sum_{\{x_1^M,\,x_2^M\}} u^M(x_1^M,\,x_2^M)\,\,\, \Rightarrow \,\,\,
u^M(x_1^M,\,x_2^M) = \frac{1}{M^2}\, \!\!\!\! \sum_{\stackrel{\{x_1,\,x_2\}}{\in\{x_1^M,\,x_2^M\}}} \!\!\!\! u(x_1,\,x_2)\,,
\end{equation}
where variables with superscript $M$ are defined in the $M\times
M$ lattice and in the second equality we are summing over all
Planck-scale lattice sites inside a single $M\times M$ cell
labelled by $(x_1^M,\,x_2^M)$. This sum computes the fraction of
populated sites in the coarse-grained cell. Finally, the sum over
$u^M$ at cell location $(x_1^M,x_2^M)$ becomes
\begin{equation}
\sum_{u^M} e^{-f\, u^M} = 1 + \binom{M^2}{1} e^{-f \frac{1}{M^2}} + \binom{M^2}{2} e^{-f \frac{2}{M^2}} + \ldots +
\binom{M^2}{M^2} e^{-f\frac{M^2}{M^2}} = \left( 1 + e^{\frac{-f}{M^2}} \right)^{M^2}.
\end{equation}
The factors in front of each exponential count how many ways a given value of $u^M$ in the coarse-grained lattice can be attained
in terms of the Planck scale lattice.  The complete partition function becomes
\begin{equation}
Z^{M\times M}  =  \prod_{x_1,x_2 \in M \mathbb{Z}} \left( 1 + e^{-\frac{\beta}{M^2} (\frac{x_1^2+x_2^2}{2} - \mu )} \right)^{M^2}
=  \frac{M^2}{\beta} F_2(e^{\frac{\beta\mu}{M^2}})\,.
 \Label{eq:MMpartfn}
\end{equation}
Thus, coarse-graining the phase space lattice size rescales the
temperature, i.e. $T\to T\,M^2$. This derivation reproduces the
semiclassical computation if we identify the temperature $\beta$
in \eqref{partfn} as the rescaled one.    We could view the
computation \eqref{eq:MMpartfn} as a derivation for the entropy
formula \eqref{localent}.

\section{The microscopic origin of macroscopic moments}

In the previous section we computed the partition function over
the $\frac{1}{2}$-BPS sector, fixing the energy $E$ and the
five-form flux $N$ using Lagrange multipliers $\beta,\, \mu$.  In
this section, we shall analyse other interesting ensembles in
which different charges are fixed in the same way.

\subsection{Fixing the integrable charges}

The typical states of fixed energy have a spectrum of the charges
$M_k$ defined in (\ref{Mk}) that is fixed by the universal $u$
function that describes their classical limit \cite{vijayjoan}.
However, it is clear that classical spacetimes with different
angular moments are easily constructed by varying the radial
dependence of the $u$ function.  Hence it is of interest to ask
what spacetime gives the universal classical description of the
typical state with fixed moments $M_k$.   We can also use the
measure (\ref{measure}) on half-BPS spacetimes to find out how
many classically indistinguishable microstates are described by
the same metric. We also see that classical spacetimes with
atypical macroscopic moments are exponentially unlikely in the
space of all states.

A natural way to identify typical states carrying fixed values of
these charges is to analyze ensembles of states that generalize
the grand-canonical one by introducing as many chemical potentials
$\mu_k$ as conserved charges $M_k$ we want to fix. As usual, we
will fix the values of these charges by requiring the expected
values to match our desired value, a step that determines the set
of chemical potentials.

While it is technically difficult to achieve this description in
the basis of states provided by the Young tableaux, it is very
natural in the coherent state basis. Indeed, the semiclassical
expressions \eqref{Mk} are linear in the phase space density
$u_0$, and so they are straightforward to integrate in the path
integral that defines the semiclassical partition function:
\begin{equation}
Z = \int \mathcal{D} u(x_1,x_2) \, \mu(u)  \, e^{-\beta (H(u) - \mu N(u) )}\,e^{-\mu_k\,M_k} \equiv
 \mathcal{D} u(x_1,x_2) \, \mu(u)\, e^{-\sum_{i=0}^N \mu_i\,M_i}\,.
\end{equation}
Above we used $M_0=N$ and $M_1=E$ and relabelled their chemical
potentials to reduce the notation. Using the measure
\eqref{measure} and evaluating the partition function by the
method of saddlepoints, its logarithm becomes
\begin{equation}
\ln Z^{M_i} = \int_0^{\infty} dx \frac{\sum_i i \mu_i
x^i}{e^{\sum_i \mu_i x^i} + 1}\,,
\end{equation}
which is a natural extension of the usual Fermi--Dirac statistics.
The value of the phase space density $u$ is fixed by the
saddlepoint:
\begin{equation}
  u = \frac{1}{1+ e^{-\sum_{i=0}^N \mu_i\,M_i}}\,.
\end{equation}
In this ensemble it is trivial to find an expression for the
expected moments; it is simply
\begin{equation}
\bar{M}_n = -\partial_{\mu_n} \ln Z.
\end{equation}

\subsection{Moments in the $(x_1,x_2)$ plane}

The eigenstates of the $M_k$ (\ref{Mk}) that were considered above
have phase space densities that are rotationally invariant in the
phase space and thus correspond to $u$ functions that are
rotationally invariant in the $x_1,\,x_2$ plane.  In particular,
any half-BPS state whose single particle phase space density is
not rotationally invariant cannot be an eigenstate of the
Hamiltonian.\footnote{For a harmonic oscillator, time translations
act as rotations of the phase space.  Energy eigenstates are
stationary, and so have time-translation invariant phase space
densities, which are thus rotationally invariant in phase space.}
Density functions $u$ with angular dependence can be efficiently
parameterized by expanding them in a Fourier series\footnote{To
keep the expression explicitly real, we use a cosine and sine
series rather than an exponential series.}:
\begin{equation}
u(r,\phi) = u_0(r) + \sum_{p=1}^{\infty} u_p^c(r) \cos p\phi + \sum_{p=1}^{\infty} u_p^s(r) \sin p\phi\,. \Label{fourier}
\end{equation}

To work out the kinds of states which give rise to  $u$ functions
with $u^c_p,u^s_p \neq 0$ we work  with the Young tableaux basis
of states,  and consider  superpositions of the form:
\begin{equation}
\Psi(\vec{q}) = \sum_w c_w \Psi^{\mathcal{F}^w}(\vec{q})\,. \Label{eq:state}
\end{equation}
Above, $w$ indexes the basis states in the superposition, each
given as a Slater determinant of single particle states as
$\Psi^{\mathcal{F}^w}(\vec{q}) = \frac{1}{\sqrt{N!}} \det
(\psi_{f^w_i}(q_j))$, and $\mathcal{F}^w$ is the set of excitation
numbers characterizing the basis state
$\Psi^{\mathcal{F}^w}(\vec{q})$. The normalization is such that
$\sum_w |c_w|^2 = 1$. The states \eqref{eq:state} are $N$ particle
states, whereas the phase space density $W$ which is mapped onto
the spacetime data $u$ (\ref{uWcorrespond})  is a one particle
density.

The method for calculating the  one-particle Wigner density
function is given in \cite{vijayjoan} (also see Appendix A). Using
this, it is easy to show that because the wavefunctions
$\psi_{f_i}(q_j)$ are orthogonal, the only interference terms
between various pieces of the superposition (\ref{eq:state}) that
survive involve states $w$ and $w'$ having $N-1$ of the fermion
excitations in $\mathcal{F}^w$ and $\mathcal{F}^{w'}$ equal. Let
us denote by $\mathcal{C}$ the set of pairs of states of this
kind. For each pair define the differing excitation $a_{(w,w')}$
by $\{a_{(w,w')}\} = \mathcal{F}^w \backslash \mathcal{F}^{w'}$.
With this definition we can now compute the one particle Wigner
distribution corresponding to (\ref{eq:state})
\begin{eqnarray}
W(q,p) & = & \int \frac{dy}{\hbar} \rho_1(q-y,q+y)
e^{\frac{2ipy}{\hbar}} = \frac{2\,e^{-\frac{r^2}{\hbar}}}{\hbar}
\sum_{w} |c_w|^2 \sum_{f \in \mathcal{F}^w} (-1)^f L_f(\frac{2r^2}{\hbar}) \nonumber \\
& + & \frac{2\,e^{-\frac{r^2}{\hbar}}}{\hbar}
\sum_{\substack{(w,w')\in \mathcal{C} \\ m<n}} (-1)^{N_m+N_n}
(-1)^m 2^{\frac{n-m}{2}+1} \sqrt{\frac{m!}{n!}} \left(
\frac{r}{\sqrt{\hbar}} \right)^{n-m}
L_m^{n-m} (\frac{2r^2}{\hbar}) \nonumber  \\
& & \times \left( \textrm{Re} \left\{ c_{w'} c^*_w \right\} \cos
(n-m)\phi + \textrm{Im} \left\{ c_{w'} c^*_w \right\} \sin (n-m)
\phi \right). \Label{eq:wigner}
\end{eqnarray}
The shorthand notation $n = a_{(w,w')}$ and $m = a_{(w',w)}$ was
used above for the sake of clarity. We also defined $N_n$ to index
the number of the differing excitation, i.e. if $n = a_{(w,w')} =
f_i^w \in \mathcal{F}^w$, then $N_n \equiv i$.

Eq. \eqref{eq:wigner} illustrates how interference between Young
tableau basis states in the superposition (\ref{eq:state}) is
responsible for breaking the $\U(1)$ invariance in the $y=0$
plane, i.e. all $\phi$ dependence in $u(r,\phi)$ is due to
interference between basis states. To quantify this, we match the
Fourier components of (\ref{fourier}) with those of
(\ref{eq:wigner}) giving
\begin{eqnarray}
u_0(r) & = & 2 e^{-\frac{r^2}{\hbar}} \sum_{w=1}^M |c_w|^2 \sum_{f \in \mathcal{F}^w}
(-1)^f L_f(\frac{2r^2}{\hbar}), \nonumber \\
u_p^c(r) & = & 2 e^{-\frac{r^2}{\hbar}} \sum_{\substack{(w,w')\in \mathcal{C} \\ n-m=p}} (-1)^{N_m+N_n} (-1)^m 2^{\frac{p}{2}+1}
\sqrt{\frac{m!}{n!}} L_m^p (\frac{2r^2}{\hbar}) \left(
\frac{r}{\sqrt{\hbar}} \right)^p \textrm{Re} (c_{w'} c^*_w), \\
u_p^s(r) & = & 2 e^{-\frac{r^2}{\hbar}} \sum_{\substack{(w,w')\in \mathcal{C} \\ n-m=p}} (-1)^{N_m+N_n} (-1)^m 2^{\frac{p}{2}+1}
\sqrt{\frac{m!}{n!}} L_m^p (\frac{2r^2}{\hbar}) \left( \frac{r}{\sqrt{\hbar}} \right)^p \textrm{Im} (c_{w'} c^*_w).
\end{eqnarray}
This shows that we have a simple relation between the fourier
modes of $u$ and the states in the superposition: {\it a given
mode $u_p^{c,s}(r)$ arises from interference between basis states
that share $(N-1)$ of their excitations, but differ by $p$ units
in the remaining excitation}. Thus the energy difference between
the two states is $p\hbar$.

While the above shows that breaking of $U(1)$ invariance in the
$y=0$ plane necessarily arises from interference in a Young
tableau basis, we should also recall that the geometric
description is only valid if we have an approximate eigenstate of
the metric operator.  We have seen in Sec.~4 that any given Wigner
distribution can be constructed out of coherent states for the
fermions and that such a state will be a eigenstate of the metric
operator.   Such states can also be written as complicated
superpositions in a Young tableau basis.

We would like to identify what typical states correspond to which
violations, and count how many of them there are. This is again
difficult to achieve in the Young tableaux basis, but it is
straightforward in the coherent state basis by introducing the
appropriate Lagrange multipliers fixing the appropriate charges.

To quantify the angular moments, define the quantities
\begin{equation}
\mathcal{U}_p^c \equiv \int \frac{d^2r}{2\pi} u(r,\phi) \cos p\phi = \frac{1}{2} \int_0^{\infty} dr \ r u_p^c(r)\,, \quad p > 0,
\Label{Ucharge}
\end{equation}
with a similar expression for $\mathcal{U}_p^s$. Note that for
$p=0$ the corresponding quantity is just the flux of the five-form
field, $\int_0^{\infty} dr\, r\, u_0(r) = N$.

With this definition we can construct the ensemble by introducing
Lagrange multipliers $\alpha_p^c$ and $\alpha_p^s$ as
\begin{equation}
e^{-\beta(E-\mu N) + \sum_{p=1}^{\infty} (\alpha_p^c \mathcal{U}_p^c + \alpha_p^s \mathcal{U}_p^s)},
\end{equation}
from which we can again compute the partition function to be
\begin{equation}
Z_{\mathcal{U}} = \prod_{x_1,x_2} \left( 1 + e^{ -\beta(\frac{r^2}{2}-\mu) + \sum_{p=1}^{\infty} (\alpha_p^c \cos p\phi +
\alpha_p^s \sin p\phi)} \right).
\end{equation}
The logarithm can again be computed
\begin{eqnarray}
\ln Z_{\mathcal{U}} & = & \frac{1}{2\pi \beta} \int_0^{\infty} \int_0^{2\pi} \frac{x}{e^{-\beta\mu - \sum_{p=1}^{\infty}
(\alpha_p^c \cos p\phi + \alpha_p^s \sin
p\phi)} e^x+1} \nonumber \\
& = & \frac{1}{2\pi \beta} \int_0^{2\pi} F_2(e^{\beta \mu + \sum_{p=1}^{\infty} (\alpha_p^c \cos p\phi + \alpha_p^s \sin
p\phi)})\, d\phi. \Label{eq:z-moments}
\end{eqnarray}
Thus we see that fixing the fourier modes $\mathcal{U}_p$ is very
natural in this language, though the integrals are again difficult
to compute and we are no longer working with Fermi--Dirac
statistics.

\paragraph{Exponential suppression of moments: }
We shall now show that the number of states with non-zero moments
$\mathcal{U}^{c,s}$ is exponentially smaller than the number of
states with $\mathcal{U}^{c,s} = 0$. To show this, we shall fix
one of the moments, chosen to be $\mathcal{U}_p^c \equiv
\mathcal{U}$, and consider the density matrix of this ensemble:
$\hat{\rho} = \frac{1}{Z} e^{-\beta(\hat{E} - \mu \hat{N}) +
\alpha\hat{\mathcal{U}}}$. From this it follows that the entropy
is
\begin{equation}
S_{\alpha} = -\langle \ln \hat{\rho} \rangle = \ln Z + \beta E - \beta \mu N - \alpha \mathcal{U}, \Label{entropy-with-violation}
\end{equation}
where the partition funtion is given by (\ref{eq:z-moments}), and
$E,N$ and $\mathcal{U}$ are determined in the standard way by
differentiating the partition function with respect to the
Lagrange multipliers:
\begin{eqnarray}
E & = & T^2 \partial_T \ln Z + T\mu \partial_{\mu} \ln Z = \frac{1}{2\pi \beta^2} \int_0^{2\pi} F_2(e^{\beta \mu +
\alpha \cos p\phi})\, d\phi, \nonumber \\
N & = & T \partial_{\mu} \ln Z = \frac{1}{2\pi \beta} \int_0^{2\pi} F_1(e^{\beta \mu + \alpha \cos p\phi})\, d\phi, \nonumber \\
\mathcal{U} & = & \partial_{\alpha} \ln Z = \frac{1}{2\pi \beta} \int_0^{2\pi} \cos p\phi \ F_1(e^{\beta \mu + \alpha \cos
p\phi})\, d\phi. \Label{eq:thermo}
\end{eqnarray}

Analyzing these relations exactly is complicated and we need to
specify a regime of parameters in which to work.  To produce
macroscopic angular  moment, we choose to work with
states of high energy or equivalently with a low density of
fermions. This corresponds to working in the limit $\mu \to
-\infty$, such that $e^{\beta \mu \pm \alpha} \to 0$.  In this
limit the Fermi--Dirac functions can be approximated by
$F_n(e^{\beta \mu \pm \alpha}) \approx e^{\beta \mu \pm \alpha}$,
and the relations (\ref{eq:thermo}) simplify considerably.

The remaining integrals in (\ref{eq:thermo}) can be computed using the calculus of residues, and after a short computation they
yield
\begin{eqnarray}
\ln Z & \approx & N \approx \beta E \approx \frac{e^{\beta\mu}}{\beta} \sum_{n=0}^{\infty} \frac{1}{n!^2}
\left(\frac{\alpha}{2}\right)^{2n}, \nonumber \\
\mathcal{U} & \approx & \frac{e^{\beta\mu}}{\beta} \sum_{n=0}^{\infty} \frac{1}{n!(n+1)!}\left(\frac{\alpha}{2}\right)^{2n+1}.
\Label{eq:thermo-sum}
\end{eqnarray}
To invert these relations let us choose the Lagrange multiplier
$\alpha$ to be very small, which will correspond to choosing the
moment to be small, but still macroscopic and measurable.
Truncating the sums to their first two terms, we can invert the
relations above to give
\begin{eqnarray}
\beta =  \frac{N}{E} & \equiv & \frac{1}{xN}, \nonumber \\
\mu =  -\frac{E}{N} \ln [\frac{E}{N^2}(1+\frac{\mathcal{U}}{N}^2)] & \equiv & -xN \ln [x(1+a^2)], \nonumber \\
\alpha =  2\frac{\mathcal{U}}{N} & \equiv &  2a,
\end{eqnarray}
where we have defined the scalings $E = xN^2$ and $\mathcal{U} =
aN$. Note that the assumption of high energy requires large $x$,
which immediately results in $\beta \mu \ll -1$, which is
consistent with our assumption. Also, the angular moment is still
measurable, since it scales linearly with $N$, with $\alpha$ small
but fixed.

Inserting these relations into (\ref{entropy-with-violation}) we get
\begin{equation}
S_{\alpha} = N(\ln x + 2) - N(\frac{2a^2}{1+a^2} - \ln (1+a^2)),
\end{equation}
which shows that the number of states with macroscopic moments of
size $aN$ is exponentially suppressed from the rotationally
invariant case:
\begin{equation}
\frac{e^{S_a}}{e^{S_{a=0}}} = e^{ - N(\frac{2a^2}{1+a^2} - \ln (1+a^2))}.
\Label{expsmall}
\end{equation}

\paragraph{Variances: }
Since fixing the macroscopic moment exponentially reduces the
number of available states, it is necessary to check that the
fluctuations in the previous computation are not so large as to
render the computation invalid. The spread in the expectation
value of $\mathcal{U}$ is given by the standard deviation to mean
ratio, which can be computed to give
\begin{equation}
\frac{\sigma(\mathcal{U})}{\langle \mathcal{U} \rangle} = \sqrt{\frac{ \langle \mathcal{U}^2 \rangle - \langle \mathcal{U}
\rangle^2}{\langle \mathcal{U} \rangle^2}} = \sqrt{\frac{ \frac{\partial_{\alpha}^2 Z}{Z} - \left(\frac{\partial_{\alpha}
Z}{Z}\right)^2} {\left( \frac{\partial_{\alpha}^2 Z}{Z} \right)^2 }} \approx \frac{1}{a\sqrt{2N}},
\end{equation}
where the computation was carried out in the previous high energy
regime, using the relations (\ref{eq:thermo-sum}). This vanishes
as $N \to \infty$, and therefore this ensemble is statistically
valid.

\paragraph{Measurability of interference: }
Just as eigenstates of the Hamiltonian generate non-trivial higher
conserved charges at infinity, we can ask whether the presence of
angular moments in the $(x_1,\,x_2)$ plane, arising from
interference between Young tableau components in the superposition
(\ref{eq:state}),
 has any characteristic effect asymptotically.
In \cite{us} we analysed how the integrable charges $M_k$ associated to Young tableau states appear at
different orders in an asymptotic expansion of $u$, the function
determining the metric. For the question at hand the same method
turns out to be very cumbersome, as the Young tableaux basis is
not a convenient one to use when analysing the moments in the $x_1,x_2$ plane.
For completeness, we still include the asymptotic expansion analogous
to (\ref{eq:wignerlimit}) for the superposition state
(\ref{eq:state}):
\begin{eqnarray}
u(\rho,\theta,\varphi) & = & 2\cos^2 \theta \sum_{l=0}^{\infty} \frac{\hbar^{l+1} \sum_{w} |c_w|^2 2^l \langle
M_l^{\mathcal{F}^w} \rangle}{\rho^{2l+s}} (-1)^l(l+1)\, \,
{_2F_1}(-l,l+2,1;\sin^2 \theta) \nonumber \\
& + & 4 \cos^2 \theta \sum_{l=1}^{\infty}
\frac{(-1)^l\hbar^{\frac{l}{2}+1}}{\rho^{l+2}}
\sum_{i=[\frac{l}{2}] +1}^l (-1)^i (i+1) \sin^{2i-l} \theta
\sum_{p=0}^{[ i-\frac{l+1}{2}]} 2^{\frac{l}{2}+p-i} \nonumber
\\ &&
\times \binom{i}{2i-l} \binom{2i-l}{p} \sum_{\substack{(w,w')\in
\mathcal{C}
\\ n-m=2(i-p)-l}} (-1)^{N_m+N_n} \sqrt{\frac{n!}{m!}} \textrm{Re} (c_m^* c_n e^{i(n-m)\phi}) \nonumber \\ & &
\times \left(\frac{d}{da}\right)^{l-i+p} \left[ a^{i-p} (a-1)^m
\right]_{a=2}. \Label{horror}
\end{eqnarray}
The first row of this expansion is a weighted sum of the single
state contributions of (\ref{eq:wignerlimit}), while the last
three rows arise from interference between the states in the
superposition. As one might imagine, analysing interference
directly from this expression would be very complicated, but we
want to point out that all odd powers of $\frac{1}{\rho}$ in the
expansion come from the interference terms. Thus, if an asymptotic
observer measures such a moment, she could infere that the underlying quantum
state {\it necessarily} was in a linear superposition of Young
tableaux.

To examine whether these moments are semiclassically measurable, it is more
convenient to work with the coherent state picture of section
(\ref{sec:coherent}).  Using coherent states we can engineer any
moments $\mathcal{U}^{c,s}_p$ we wish, by distributing the
fermions in a suitable way. As an example, let us pick a very
specific configuration with exactly one $U(1)$ violating moment
turned on:
\begin{equation}
u(r,\phi) = \frac{1}{2} \Theta(\sqrt{2}R^2-r) (1+2a\cos \phi), \Label{examplemoment}
\end{equation}
where $\Theta$ is the Heaviside step function, $a$ is a parameter
controlling the size of $U(1)$-violation and $R = (2 \hbar
N)^{\frac{1}{4}}$ is the AdS radius.\footnote{The factor of
$\sqrt{2}$ in the argument of the step function is necessary to
ensure the correct normalization: $\int_{\mathbb{R}^2}
\frac{u(r,\phi)}{2\pi \hbar} d^2\vec{r} = N$.}  One sees that for
this distibution $\mathcal{U}_1^c = aN$, and therefore for small
$a$ the state is of the form analyzed earlier in this section.

An asymptotic expansion of the function $u(\rho, \theta, \phi)$
corresponding to this fermion distribution can be computed via the
technique used in \cite{us}, and the first non-trivial terms turn
out to be
\begin{equation}
u(\rho,\theta,\phi) = \frac{\cos^2 \theta}{\rho^2} \hbar N + \frac{2^{\frac{5}{2}}a}{3} \frac{\cos^2 \theta \sin \theta \cos
\phi}{\rho^3} (\hbar N)^{\frac{3}{2}} + \mathcal{O}(\rho^{-4})\,.
\end{equation}
The $(\hbar N)^{3/2}$ dependence of the first odd $\rho$ negative
power illustrates the finiteness of this moment in the
semiclassical limit:
\begin{equation}
N \to \infty, \quad \hbar \to 0, \quad \textrm{ such that } \hbar N = \textrm{ fixed}. \Label{semiclass}
\end{equation}
Thus, the interference between the Young tableau components of
some states is measurable by an asymptotic observer, even though
for a typical state these effects are highly suppressed as shown
in (\ref{expsmall}).

\section{$ N \rightarrow 1$-particle information loss}
\llabel{infloss}

Our work above has emphasized that individual quantum states do
not generically correspond to individual geometries.  In
particular, we have shown how a geometry arising from a specified
function $u(x_1,x_2)$ as in (\ref{udef}) can be the effective
coarse-grained description of many quantum states.  As noted in
section \ref{coarsegrained}, the function $u$ determines only the
expectation value of the (coarse-grained) operator $\tilde u$.
Furthermore, this function encodes only information present in the
one-particle phase space density \cite{vijayjoan, us} derived by
projection of the full quantum state.  Below, we first
characterize more precisely when this projection loses
information.  It will turn out that in the semiclassical limit,
the 1-particle projection loses information if and only if the
description as an LLM geometry is singular.  We then characterize
the operators which probe {\it beyond} this one-particle
projection from several points of view. (See \cite{vijayjoan} for
a discussion of related issues.)

\subsection{Singularities and loss of information}
\llabel{sec:density}

In Sec.~\ref{superstar-example} we gave four examples of states
that give rise to single particle phase space distributions that
reproduce the metric of the `superstar' extremal black hole. Two
of these states, ``Square + Square" (\ref{squaresquare}) and
``Triangle + Triangle" (\ref{triangletriangle}), have identical
single particle distributions even though they are very different
as $N$-particle states.  In this case, the ``Square + Square''
state does not have a description as a single geometry since it is
not an approximate eigenstate of the metric operator.  But it is
easy to devise examples of multiple states, all of which have good
descriptions as a single coarse-grained geometry, and which have
identical single particle phase space distributions.  The projection to the single
particle phase space has lost information about the underlying
$N$-particle configurations.

To see when this situation may occur, suppose that the support of
the function $u$ on the $(x_1,x_2)$ plane has an area  $2 \pi
\hbar k$. As in section \ref{sec:coherent}, we envision the
fermions as coherent states inhabiting a lattice, and identify the
support of $u$ with lattice sites $\{ \lambda_1, \ldots, \lambda_k
\} = {\rm supp}\, u$.  Also, because there are $N$ fermions, we
must have
\begin{equation}
\int \frac{dx_1 dx_2}{2\pi\hbar} \, u = N \, . \Label{norm}
\end{equation}
Thus, the wavefunction is a sum of $N$-particle components, each
of which is of the form $|\lambda_{i_1}, \ldots,
\lambda_{i_N}\rangle$, where $i_s \in {\rm supp}\, u ~~ \forall
s$. There are $\binom{k}{N}$ such $N$-particle components, so the
superposition coefficients required to specify the general state
involve $2\binom{k}{N}-2$ real numbers (where we subtract $2$ to
account for the normalization and the unphysical overall phase).
On the other hand, since the Wigner distribution is
nonzero at $k$ lattice sites, it contains $k-1$ real numbers
(after accounting for the normalization
(\ref{norm})).\footnote{Note that the associated 1-particle
density matrix is rank $k$, and, up to $\hbar$ corrections,  is
diagonal for coherent states based on lattice sites that give rise
to approximate metric eigenstates.  Thus it also only contains
$k-1$ real numbers in the semiclassical limit.}    Therefore, when
$k>N$, the 1-particle projection simply does not contain enough data to
determine the underlying $N$-particle state.

In any situation where the effective phase space distribution has
support on more than $N$ lattice sites, the normalization
(\ref{norm}) will force the distribution to take values that are
neither $0$ nor $1$ in at least some locations.     In that case,
following the correspondence (\ref{uWcorrespond}), the spacetime
geometry is necessarily singular.  Since the density takes values other than $0$ or $1$, we see that the fine-grained metric also fluctuates; our state is not an approximate eigenstate of the fine-grained $\hat u$. In this sense then, singularities in the
half-BPS sector of $\ads{5}$ gravity arise precisely when a single
classical geometry is unable to encode some data concerning the
underlying quantum mechanical configuration space. This is satisfactory because any attempt to do physics in these singular spacetimes forces us to specify boundary conditions on the singularity. As mentioned in \cite{vijayjoan}, it would be very interesting to understand how these boundary conditions emerge from the effect of coarse-graining of wave functions applied to correlation functions in the
gauge theory.

\subsection{Operators probing $N$-particle structure}

\paragraph{Second-quantized formalism: }
The second-quantized formalism of section \ref{2nd} makes it
particularly easy to exhibit operators sensitive to more than
1-particle structure.  We simply string together
creation and annihilation operators corresponding to different
sites on the LLM plane. One example is
\begin{equation}
\hat{\mathcal{O}}_{10} = b^\dagger (1) b (1) b^\dagger (0) b (0)\,
,
\end{equation}
where $b(0)$ stands for the second-quantized operator that
destroys a particle in the $0^{{\rm th}}$ level in the energy
eigenbasis. Then:
\begin{equation}
\begin{array}{lcl}
\hat{\mathcal{O}}_{10} | \textrm{`triangle + triangle'} \rangle &
= & 0
\\
\hat{\mathcal{O}}_{10} | \textrm{`square + square'} \rangle & = &
\frac{1}{\sqrt{2}} |0,1,\ldots, N-1 \rangle \,
\end{array}
\end{equation}
where the two states above are specified in
(\ref{squaresquare},\ref{triangletriangle}).   Here the `square +
square' state does not have a good description as a single
geometry, but we could have equally given an example where the
$N$-particle distinguished two states that have both have good
descriptions in an identical geometry.

\paragraph{Wigner's formalism: }
Recall that the expectation value of a 1-particle
operator $\hat{\mathcal{O}}$ in a state represented by the
1-particle reduced Wigner distribution $W_1 (\alpha, \alpha^*)$ is
\begin{equation}
\langle \hat{\mathcal{O}} \rangle = \frac{\hbar}{\pi}
\int_{\mathbb{C}} d^2 \alpha f_{\mathcal{O}} (\alpha, \alpha^*)
W_1(\alpha, \alpha^*) \, ,
\end{equation}
where $f_{\mathcal{O}} (\alpha, \alpha^*)$ is the classical phase
space distribution corresponding to $\hat{\mathcal{O}}$. The
formalism of Wigner distributions extends to calculating
expectation values of operators sensitive to multi-particle
structure. The $N$-particle Wigner distribution is given by:
\begin{equation}
  W_N(\vec{q},\vec{p})=\hbar^{-N}
 \int_{-\infty}^{\infty} d\vec{y} \langle \vec{q}-\vec{y}|
\hat{\rho}_N | \vec{q}+   \vec{y} \rangle \,e^{2i \vec{p} \cdot
\vec{y} / \hbar} \, .
\end{equation}
From this, the 1-particle reduced Wigner distribution is recovered
by tracing over $N-1$ particles:
\begin{equation}
W(\alpha, \alpha^*) = \frac{N\,\hbar^{N-1}}{\pi^{N-1}} \int d^2
\alpha_2 \ldots d^2 \alpha_N W_N(\vec{\alpha},\vec{\alpha}^*)\, .
\end{equation}
In analogy to the  1-particle case, expectation values
of operators in states characterized by $W_N$ are given by:
\begin{equation}
\langle \hat{\mathcal{O}} \rangle = \frac{\hbar^N}{\pi^N}
\int_{\mathbb{C}^N} d^{2N} \vec{\alpha} f_{\mathcal{O}}
(\vec{\alpha}, \vec{\alpha}^*) W_N(\alpha, \alpha^*) \, .
\end{equation}
Note that $f_{\mathcal{O}} (\vec{\alpha}, \vec{\alpha}^*)$ is
necessarily symmetric in the $\alpha_i$'s.

From the above, it is clear that operators admitting a
representation $\mathcal{O}(\vec{\alpha},\vec{\alpha}^*)=\sum_i
f(\alpha_i,\alpha_i^*)$ can be completely evaluated in the
1-particle reduced Wigner distribution:
\begin{eqnarray}
\langle \hat{\mathcal{O}} \rangle & = &
      \frac{\hbar^N}{\pi^N}
      \int_{\mathbb{C}^N} d^{2N} \vec{\alpha} \textrm{ } W_N (\vec{\alpha}, \vec{\alpha}^*) \sum_i f(\alpha_i,\alpha_i^*) \nonumber \\
& = & \frac{\hbar}{\pi} \sum_i \int_{\mathbb{C}} d^2 \alpha_i
f(\alpha_i,\alpha_i^*) \left( \frac{\hbar^{N-1}}{\pi^{N-1}}
\int_{\mathbb{C}^{N-1}} \prod_{j \not{=} i} d^2 \alpha_j
            W_N (\vec{\alpha}, \vec{\alpha}^*) \right) \nonumber \\
& = & \frac{\hbar}{\pi} \sum_i \int_{\mathbb{C}} d^2 \alpha_i f(\alpha_i,\alpha_i^*) \left( \frac{1}{N} W_1(\alpha_i, \alpha_i^*) \right) \nonumber \\
& = & \frac{\hbar}{\pi} \int_{\mathbb{C}} d^2 \alpha f(\alpha,
\alpha^*) W_1(\alpha, \alpha^*) \, .\nonumber
\end{eqnarray}
On the other hand, any operator whose classical density is not of
the form $\sum_i f(\alpha_i,\alpha_i^*)$ (i.e. contains a monomial
in $\alpha$'s of mixed indices) will generically probe the
$N$-particle structure. The space of 1-particle operators with
non-vanishing expectation values admits a basis given by the
moments $M_n$ (\ref{higherham}) with classical phase space
densities
\begin{equation}
\mathcal{O}_n = \sum_i (\alpha_i\alpha_i^*)^n = M_n \, .
\end{equation}
This is intuitive as the knowledge of all the $M_n$'s is in
principle necessary and sufficient for specifying $W_1$. Any
operator which is not a linear combination of the $M_n$'s will in
general probe the $N$-particle information.

\paragraph{$N$-particle operators in full AdS/CFT: }

The recognition of 1-particle operators as the higher Hamiltonians
of the fermionic system afford a clean characterization of
operators probing the $N$-particle structure in the full CFT. In
particular, any operator which may not be written as a linear
combination of $\hat{\mathcal{O}}_n = \textrm{Tr}\, (X^{\dagger n}
X^n)$ will belong to this family. $R$-charge conservation and
gauge invariance translates this requirement into the following:
operators probing the $N$-particle structure are multi-trace
operators. Not surprisingly, multi-trace operators correspond to
multi-particle states on the gravity side \cite{aharony}. They
have been investigated in \cite{witten, berkooz}. A simple example
of such an operator is
\begin{equation}
\textrm{Tr}\, (X^\dagger X) \, \textrm{Tr}\, (X^\dagger X) -
\textrm{Tr}\, (X^{\dagger 2} X^2) \, ,
\end{equation}
where the second term removes the 1-particle component. Its
classical phase space distribution is given by
\begin{equation}
\{ \, |\alpha_1|^2 |\alpha_2|^2 \, \} \, ,
\end{equation}
where $\{ \ldots \}$ denotes complete symmetrization over
$\alpha_1, \ldots, \alpha_N$.

\section{Discussion}
\llabel{discussion}

Our definition of the metric operator $\hat u(x_1,x_2)$ was based
on the use of standard coherent states, having equal dispersions
in $x_1$ and $x_2$.  Similarly, our notion of coarse-graining was
based on Gaussian smearing kernels having equal $x_1$- and
$x_2$-dispersions.  This choice was made for the sake of
simplicity and, furthermore, for an observer who makes
sufficiently rough measurements (i.e., who is sufficiently
semi-classical) of a fixed state, the detailed shape of
$\hbar$-sized phase space cells is irrelevant.   However, consider
a given ``semi-classical'' observer who is able to measure
energies with some fixed accuracy  $\delta E$.   Suppose that this
observer studies one of the above coherent states centered on a
point with $r^2 = x_1^2 + x_2^2$.  This state  has an energy
uncertainty $\Delta E \sim r \delta r / \hbar  \sim r /
\sqrt{\hbar}$.  Clearly for large $r$ we have $\Delta E \gg \delta
E$ and the observer's measurements will cause this state to
quickly decohere.

Such measurements are not well-described by our formalism above.
Instead, such observers require an analogous formalism based on
{\it squeezed} coherent states, where the squeezing reduces the
radial dispersion to achieve  $\delta r \sim  \hbar/r$ at the
expense of increasing the angular dispersion to the level $\delta
\theta \sim 1$, such that $r \delta r \delta \theta \sim \hbar$.
They also require a correspondingly squeezed notion of coarse
graining.  It would be very interesting to study such a squeezed
formalism in detail, but we have chosen the simpler (unsqueezed)
formalism for our work above.  It would also be interesting to
analyze bulk semi-classical observers from first principles to
verify that fixed energy resolution $\delta E$ (as opposed to a
$\delta E$ which changes with $E$) is in fact an appropriate
description of their measurements.

\section*{Acknowledgements}
We thank Jan de Boer,  Tamaz Brelidze, Veronika Hubeny, Vishnu
Jejjala, Thomas Levi, Rob Myers, Mukund Rangamani, Simon Ross,
Slava Rychkov and Jung-Tay Yee  for useful discussions. B.C.,
K.L., V.B.\ and J.S.\ were supported in part by the DOE under
grant DE-FG02-95ER40893. K.L.\ was supported in part by the Netter
Fellowship from the University of Pennsylvania, and by a
fellowship from the Academy of Finland.   B.C. was supported in
part by a Dissertation Completion Fellowship from the University
of Pennsylvania.    V.B. thanks the theoretical physics group at
Berkeley and the organizers of the Sowers Workshop in Theoretical
Physics at Virigina Tech for hospitality while this paper was
finished.  D.M. was supported in part by the National Science Foundation under Grant No PHY03-54978, and by funds from the University of California.
J.S. was supported in part by DOE grant DE-AC03-76SF00098 and NSF grant PHY-0098840.

\appendix
\section{Computing the Wigner distribution}
In this appendix we present the computation of the Wigner
distribution for a general state in the $N$-particle harmonic
oscillator, deriving (\ref{eq:wigner}). A general state can be
written as a superposition of basis states as (\ref{eq:state}),
and the single particle wavefunctions appearing in the Slater
determinant are given by
\begin{equation}
\psi_f(q) = (\pi \hbar)^{-\frac{1}{4}} \frac{e^{-\frac{q^2}{2\hbar}}}{\sqrt{2^f f!}} H_f(\frac{q}{\sqrt{\hbar}}),
\Label{1partpsi}
\end{equation}
where $H_n$ is a Hermite polynomial. We then compute the effective
one particle density matrix as
\begin{eqnarray}
\rho_1(x,y) & = & \int dq_2 \ldots dq_N \Psi^*(x,q_2,\ldots,q_N) \Psi(y,q_2,\ldots,q_N) \nonumber \\
& = & \int dq_2\ldots dq_N \sum_{w,\tilde{w}} c_w^* c_{\tilde{w}} \epsilon^{i_1\ldots i_N} \epsilon^{1 j_2\ldots j_N}
\epsilon^{k_1\ldots k_N} \cdot \nonumber \\ & & \cdot \epsilon^{1 l_2\ldots l_N} \psi^*_{f_{i_1}^w}(x)
\psi^*_{f_{i_2}^w}(q_{j_2}) \ldots \psi^*_{f_{i_N}^w}(q_{j_N}) \psi_{f_{k_1}^{\tilde{w}}}(y) \psi_{f_{k_2}^{\tilde{w}}}(q_{l_2})
\ldots
\psi_{f_{k_N}^{\tilde{w}}}(q_{l_N}) \nonumber \\
& = & \ldots = \sum_{w=1}^M \left( |c_w|^2 \sum_{f\in \mathcal{F}^w} \psi^*_f(x) \psi_f(y) \right) \nonumber \\
\label{densitymatrix} & & + \sum_{(w,\tilde{w})\in \mathcal{C}} (-1)^{N_{(w,\tilde{w})} + N_{(\tilde{w},w)}} c^*_w c_{\tilde{w}}
\psi^*_{a_{(w,\tilde{w})}}(x) \psi_{a_{(\tilde{w},w)}}(y).
\end{eqnarray}
Before explaining the notation above, some comments are in order.
The right hand side of (\ref{densitymatrix}) consists of two
parts. The `diagonal' terms on the upper line are contributions of
single basis states, as already computed in \cite{vijayjoan}, only
now weighted by $|c_w|^2$. The second part, the `cross terms',
represent interference between basis states. Due to the
orthogonality of the 1-particle wavefunctions (\ref{1partpsi}) it
is clear that a cross term between two basis states appears if
$(N-1)$ of the excitations in these states are the same, i.e. the
sets $\{f_1^w, \ldots, f_N^w\}$ and $\{f_1^{w'}, \ldots, f_N^{w'}
\}$ have $(N-1)$ equal elements, while the remaining elements
differ. Note that when this happens, the crossed basis states
necessarily have different energies and the superposition state is
prohibited from being a hamiltonian eigenstate.

The notation is as follows: let $\mathcal{C}$ denote the set of
pairs that give rise to interference terms as explained above.
Then for each excitation we can define the differing excitation
$a_{(w,w')}$ as $\{ a_{(w,w')} \} \in \mathcal{F}^w \backslash
\mathcal{F}^{w'}$. We also define the index of the differing
excitation such that if $a_{(w,w')} = f_i^w \in \mathcal{F}^w$,
then $N_{(w,w')} = i$.

The Wigner distribution corresponding to a density matrix
$\hat{\rho}_1$ is defined as
\begin{equation}
W(q,p) = \frac{2}{\hbar} \int_{-\infty}^{\infty} dy \, \langle q-y
| \hat{\rho}_1 | q+y \rangle \, \, e^{\frac{2ipy}{\hbar}},
\end{equation}
and using equations (\ref{1partpsi}), (\ref{densitymatrix}) and the identity
\begin{equation}
\int_{-\infty}^{\infty} dx e^{-x^2} H_m(x+y) H_n(x+z) = 2^n \sqrt{\pi} m! z^{n-m} L_m^{n-m} (-2yz), \quad m \le n,
\end{equation}
it is now straightforward to verify (\ref{eq:wigner}). Above the
$L_a^b$ are associated Laguerre polynomials.

\end{document}